\documentclass[prd,12pt,english]{revtex4}
\usepackage{amsmath,amssymb} 

\usepackage{graphicx}
\usepackage{epsfig}
\usepackage{babel}

\newcommand{\ctp}{\cos\theta}

\newcommand\nn{\nonumber}
\newcommand\ba{\begin{eqnarray}}
\newcommand\ea{\end{eqnarray}}
\newcommand\ee{\end{equation}}
\newcommand\be{\begin{equation}}

\begin{document}
\title{Polarization effects in $\bar N+N\to \pi +\ell^++\ell^-$ reaction.
General analysis and numerical estimations}

\author{G. I. Gakh}
\affiliation{\it  National Science Centre "Kharkov Institute of 
Physics and Technology", Akademicheskaya 1,  61108 Kharkov, Ukraine}

\author{A. P. Rekalo}
\affiliation{\it  National Science Centre "Kharkov Institute of 
Physics and Technology", Akademicheskaya 1,  61108 Kharkov, Ukraine}

\author{E.~Tomasi-Gustafsson}
\thanks{ Permanent address: CEA,IRFU,SPhN, Saclay, 91191 Gif-sur-Yvette Cedex, France}
\affiliation{\it Univ Paris-Sud, CNRS/IN2P3, Institut de Physique Nucl\'eaire, UMR 8608, 91405 Orsay, France}

\author{J.~Boucher}
\affiliation{\it Univ Paris-Sud, CNRS/IN2P3, Institut de Physique Nucl\'eaire, UMR 8608, 91405 Orsay, France}

\author{A. G. Gakh}
\affiliation{Kharkov National University, 61077 4 Svobody Sq., Kharkov, Ukraine}
\date{\today}

\begin{abstract}
 A general formalism is developed to calculate the cross section and the polarization observables for the reaction $\bar N+N\to \pi +\ell^++\ell^-$. The matrix element and the observables are expressed in terms of six scalar amplitudes (complex functions of three kinematical variables) which determine the reaction dynamics. The numerical predictions are given in frame of a particular model in the kinematical range accessible in the PANDA experiment at FAIR.
\end{abstract}

\maketitle

\section{Introduction}

The experimental determination of the electromagnetic form factors (FFs) of the nucleon has triggered
large experimental programs at all major facilities in the GeV energy range. They have long served as one of the testing ground for our understanding of nucleon structure ranging from the large intranuclear distances, where QCD is not applicable, up to short distances which correspond to high transferred momenta, where perturbative QCD is valid. Basically, all models of nonperturbative QCD which use effective degrees of freedom have been applied to describe the nucleon electromagnetic structure.

Most of the information on the nucleon electromagnetic FFs is obtained in elastic electron scattering on nucleons, assuming that the interaction occurs through the exchange of a virtual photon with four momentum squared $q^2$. The accessible kinematical region, in this case, is denoted as space-like (SL) region and corresponds to the range of negative $q^2$ values. The investigation of the annihilation processes $ e^++e^-\leftrightarrow \bar p+p$ allows to access positive $q^2$ values (time-like region (TL)) over the reaction threshold. Analyticity and unitarity require that FFs are real functions of $q^2$ in the SL region, while for the TL region they must be complex functions.

The interest to the measurement of nucleon FFs has been renewed by the possibility to apply the polarization transfer method, first suggested in Ref. \cite{Re67}, due to the availability of highly polarized electron beams at large intensity. Measurements at MIT, MAINZ and JLAB, using the polarization transfer and target asymmetry methods, showed that the ratio $\mu G_E/G_M $ ($\mu$ is the proton magnetic moment, $G_E(G_M)$ is the proton electric(magnetic) FF) deviates from unity, in contrast with the results derived from the Rosenbluth separation technique \cite{Pu10}. The reason of this discrepancy, probably related to radiative corrections, is still controversial (see the review \cite{CFP08}).
The individual determination of TL FFs, planned by the PANDA collaboration  \cite{PANDA}, will also shed light on this problem.

In this paper we study the annihilation reaction
\be
\bar p(p_1)+p(p_2)\to \gamma^*(q)+\pi^0(q_{\pi})\to  \ell^+(k_2)+\ell^-(k_1)+ \pi^0(q_{\pi}),
\label{eq:eq1}
\ee
(the notation of the particle four-momenta is indicated in
brackets) which will be measurable in the near future at hadron colliders, in particular at the future FAIR project \cite{FAIR}. The interest of this reaction lies in the possibility of
accessing nucleon FFs in the TL region for $q^2< 4M^2$ (where $M$ is the nucleon mass) which are unaccessible in the reactions:
$e^+e^-\leftrightarrow N\bar N$. 

Similarly to the initial state radiation, the
emission of a pion by the proton or the antiproton in the initial
state lowers the $q^2$ value of the
virtual photon in the $\bar p+p\to
e^++e^-+\pi^0 $ reaction. That way, one could reach the otherwise
unaccessible $q^2$ range below the threshold and
measure the nucleon FFs down to lower $q^2$ values. This is particularly interesting as the contribution of meson resonances is expected in that region. These data will be especially important to tune the models based on dispersion relations, as well as all analytical models which allow to connect SL and TL regions. The polarization degrees of freedom, either with polarized target or with antiproton beam polarized normally to the reaction plane, allow to access the imaginary part of the complex FFs.

The reaction (\ref{eq:eq1}) is related by crossing symmetry to the pion scattering process $\pi +N\to N+e^-+e^+$, which was first studied in Ref. \cite{MPR}. Pion scattering on nucleon and nuclei will be investigated in the resonance region by the HADES collaboration. This program will make use of the pion beam which will be available at GSI (Darmstadt) in the next future \cite{Hades}. 

In Ref. \cite{DDR} a general
expression for the cross section of the reaction (\ref{eq:eq1}) was derived and
numerical estimations were given in the kinematical conditions when
antiproton annihilates at rest. In Ref. \cite{AKTM} the formalism of the previous work was extended to moving antiproton. A larger set of the Feynman diagrams was considered, with special emphasis to the
possibility of accessing the nucleon axial FFs in the TL region \cite{Ad07}. The differential cross section in an experimental setup
where the pion is fully detected was given, with explicit dependence
on the relevant nucleon FFs.

Studies are ongoing to polarize antiproton beams and/or to install a polarized  proton target at the future FAIR facility \cite{PAX}. In this perspective, we study polarization effects in the reaction (\ref{eq:eq1}). We consider the cases when only one hadron is polarized (proton or antiproton), and when both proton and antiproton have an arbitrary polarization.  
Note that the results of the analysis of the polarization effects in
the reaction (\ref{eq:eq1})  have a general nature and are not related to a
particular reaction mechanism. The derivation of these results requires P-invariance of the hadron electromagnetic interaction, hadron
electromagnetic current conservation and the fact that the photon
has spin one.

The numerical evaluation of the differential cross section and of the polarization observables requires, however, a model calculation. Using the model of Ref. \cite{AKTM}, we calculated the following
polarization  observables: the asymmetries due to the
polarization of the proton target or antiproton beam, and the
correlation coefficients which arise from the polarization of both
proton and antiproton. These estimations are done assuming an experimental setup where the pion and the virtual photon four momenta are fully determined, in the kinematical region accessible at PANDA.

\section{General formalism}
The general structure of the differential cross section and various polarization observables for the
$\bar p+p\to\pi^0+\gamma^* \to \pi^0+\ell^++\ell^-$  reaction  ($\gamma^*$ is a virtual photon) is determined on the basis of the most general symmetry properties of the
hadron electromagnetic interaction, such as gauge invariance (the
conservation of the hadronic and leptonic electromagnetic currents)
and P-invariance (invariance with respect to space reflections). It 
does not depend on the details of the reaction mechanism. The matrix element of the
reaction (\ref{eq:eq1}) can be written in terms of the hadronic $J_{\mu}$ and leptonic
$j_{\mu}$ currents as:
\be
M=\frac{4\pi\alpha}{q^2}J_{\mu}j_{\mu}, ~j_{\mu}=\bar u(-k_2)\gamma_{\mu}u(k_1),
\label{eq:eq2}
\ee
where $q=k_1+k_2$ is the virtual photon four-momentum, $J_{\mu}$ is
the electromagnetic current describing the transition  $\bar p+p\to
\pi^0+\gamma^* $.

The general expression for the differential cross section of
the reaction considered has the standard form:
\be
d\sigma =\frac{(2\pi )^4}{4I}|M|^2d\Gamma ,
\label{eq:eq3}
\ee
where $I^2=(p_1\cdot p_2)^2-M^4, $ and
$d\Gamma $ is the phase space volume given by
\be
d\Gamma =(2\pi )^{-9}\frac{d^3k_1}{2E_1}\frac{d^3k_2}{2E_2}
\frac{d^3q_{\pi}}{2E_{\pi}}\delta^{(4)}(p_1+p_2-k_1-k_2-q_{\pi}).
\label{eq:eq4}
\ee
Here $E_1 (E_2)$ is the energy of the electron (positron) and
$E_{\pi}$ is the pion energy.

The modulus square of the matrix element can be written as
contraction of the hadronic and leptonic tensors:
\be
|M|^2=\frac{16\pi^2\alpha^2}{q^4}H_{\mu\nu}L_{\mu\nu},
\label{eq:eq5}
\ee
where the hadronic $H_{\mu\nu}$ and leptonic $L_{\mu\nu}$ tensors are defined as
\be
H_{\mu\nu}=J_{\mu}J_{\nu}^*, ~   L_{\mu\nu}=j_{\mu}j_{\nu}^*.
\label{eq:eq6}
\ee
\subsection{The matrix element}

The electromagnetic structure of hadrons, as probed by elastic and
inelastic electron scattering, is characterized by a set of
structure functions. Each of these structure functions is determined
by different combinations of the longitudinal and transverse
components of the electromagnetic current $J_{\mu}$, thus providing
different pieces of information about the hadron structure or
possible mechanisms of the reaction under consideration. Similar
formalism can be applied to the annihilation reactions for the TL region of the virtual photon.

The formalism of the structure functions is  especially convenient
for the investigation of polarization phenomena in the reaction (\ref{eq:eq1}).

Taking into account the conservation of the leptonic $j_{\mu}$ and
hadronic $J_{\mu}$ electromagnetic currents the matrix element of
the  $\bar p+p\to e^++e^-+\pi^0$ reaction can be written as
\be
M=ee_{\mu}J_{\mu}=e{\vec \epsilon}\cdot {\vec J},~
e_{\mu}=\frac{e}{q^2}j_{\mu}, ~{\vec \epsilon}=\frac{{\vec
e}\cdot {\vec k}}{k_0^2}{\vec k}-{\vec e}.
\label{eq:eq7}
\ee
The structure of the matrix element is the same as for the  $\bar
p+p\to \pi+\gamma^*$ reaction, which is the crossed channel of 
$\gamma^*+p\to p+\pi $ reaction, the electroproduction of pions on the 
nucleon. The most convenient system for the analysis of polarization effects in
this last reaction is the reaction center-of-mass system
(CMS) \cite{Ak77}. Let us use a similar approach for the investigation of the
polarization effects in the reaction under consideration. 

In the initial nucleon-antinucleon CMS we get 
the following expression for the matrix element:
\be
M=e\varphi_{\pi}^+\varphi_2^+{\cal F}\varphi_1,
\label{eq:eq8}
\ee
where $\varphi_1$ $(\varphi_2)$ is the proton (antiproton) spinor,
respectively and ${\cal F}$ is the reaction amplitude which can be written 
in different equivalent forms.

Let us introduce, for convenience and for simplifying the following
calculation of polarization observables, the orthogonal system of
the basic unit vectors ${\vec q}$, ${\vec m}$ and ${\vec n}$  which
are built from the momenta of the particles participating in the
reaction
\be
{\vec q}=\frac{{\vec k}}{|{\vec k}|}, ~{\vec n}=\frac{{\vec
k}\times {\vec p}}{|{\vec k}\times {\vec p}|}, ~{\vec m}={\vec
n}\times {\vec q},
\label{eq:eq9}
\ee
where ${\vec k}({\vec p})$ is the virtual photon (antiproton)
three-momentum. The unit vectors ${\vec q}$ and ${\vec m}$ define the
$\bar p +p\to \pi+\gamma^*$ reaction plane, and the unit vector
${\vec n}$ is perpendicular to the reaction plane.

The amplitude ${\cal F}$, which describes the dynamics of the  $\bar pp\to
\pi\gamma^*$ reaction, is determined, in general case, by six
independent amplitudes for the P-conserving hadronic interaction. This 
follows from the fact that the reaction considered is the cross
channel of the process $e+N \to e+N+\pi $, the pion electroproduction on
the nucleon. The general form of the amplitude ${\cal F}$ can be written in various equivalent forms. In the analysis of the polarization phenomena, it is
convenient to use the following general form for the amplitude ${\cal F}$:
\be
{\cal F}={\vec \epsilon}\cdot {\vec m}(f_1{\vec \sigma}\cdot {\vec
m}+f_2{\vec \sigma}\cdot {\vec q})+ {\vec \epsilon}\cdot {\vec
n}(if_3+f_4{\vec \sigma}\cdot {\vec n})+ {\vec \epsilon}\cdot {\vec
q}(f_5{\vec \sigma}\cdot {\vec m}+f_6{\vec \sigma}\cdot {\vec q}),
\label{eq:eq10}
\ee
where $f_i$ ($i=1-6$) are the scalar amplitudes in orthogonal
basis, which completely determine the reaction dynamics. These
amplitudes contain information about the dynamics of the considered reaction 
and they depend on three independent kinematical variables, for example, $s=(p_1+p_2)^2$, $q^2$ and $t=(p_1-q)^2$.

\subsection{The hadronic tensor}

Let us consider the general properties of the hadronic tensor.

Taking into account that the hadronic and leptonic tensors satisfy
the following conditions:
$H_{\mu\nu}q_{\mu}=H_{\mu\nu}q_{\nu}=L_{\mu\nu}q_{\mu}=L_{\mu\nu}q_{\nu}=0$
due to the hadronic and leptonic current
conservation, one can show that the polarization observables for the
considered reaction are determined by the space components
of the hadronic tensor only.

The hadronic tensor $H_{ij}$, ($i,~j=x,~y,~z$) can be represented as the 
sum of contributions, classified according to the polarization states of the proton-antiproton system, in the following form:
\be
H_{ij}=H_{ij}(0)+H_{ij}(\xi_1)+H_{ij}(\xi_2)+H_{ij}(\xi_1,\xi_2),
\label{eq:eq11}
\ee
where the term $H_{ij}(0)$ corresponds to unpolarized proton and antiproton, the term $H_{ij}(\xi_1)$
($H_{ij}(\xi_2)$) corresponds to the case where one particle, the proton
(antiproton), is polarized and the term $H_{ij}(\xi_1,\xi_2)$ corresponds to the case when both initial particles are polarized.

The general structure of the first term of the hadronic tensor, which
corresponds to unpolarized antiproton and proton, has the following
form
\be
H_{ij}(0)=\alpha_1q_iq_j+\alpha_2n_in_j+\alpha_3m_im_j+
\alpha_4(q_im_j+q_jm_i)+i\alpha_5(q_im_j-q_jm_i).
\label{eq:eq12}
\ee
The real structure functions $\alpha_i$, $i=1-5$, depend on three
variables $s$, $q^2$ and $t$ or $\cos\theta $, where $\theta $ is
the angle between the momenta of the antiproton and the virtual photon.
Note that the structure function $\alpha_5 $, for the crossed
(scattering) channel $e+N \to e+N+\pi $, is determined by strong
interaction effects of the final-state hadrons and vanishes for the
pole diagram contribution (Born approximation) in all
kinematical range. In our case this structure function is not zero
even in Born approximation due to the fact that hadron FFs are complex here. It  contributes 
to the cross section only in the case of polarized lepton since
the lepton tensor, in such case, contains an antisymmetrical part.
The structure functions are related to the reaction scalar
amplitudes $f_i$ and the expressions of these structure functions in
terms of the scalar amplitudes are given in the Appendix.

Let us first consider the case when only one hadron is polarized in the
initial state. 

The general structure of the hadronic tensor in
the case of polarized proton can be written as
\ba
H_{ij}(\xi_1 )&=&{\vec \xi_1}\cdot {\vec n}\Bigl (\beta_1q_iq_j+\beta_2m_im_j+
\beta_3n_in_j+\beta_4\{q,m\}_{ij}+i\beta_5[q,m]_{ij}\Bigr )
\nn\\
&&
+{\vec \xi_1}\cdot {\vec q}\Bigl (\beta_6\{q,n\}_{ij}+\beta_7\{m,n\}_{ij}+
i\beta_8[q,n]_{ij}+i\beta_9[m,n]_{ij}\Bigr ) 
\nn\\
&&
+{\vec \xi_1}\cdot {\vec m}\Bigl (\beta_{10}\{q,n\}_{ij}+\beta_{11}\{m,n\}_{ij}+
i\beta_{12}[q,n]_{ij}+i\beta_{13}[m,n]_{ij}\Bigr ),
\label{eq:eq13}
\ea
where ${\vec
\xi_1}$ is the unit vector of the proton polarization, 
$\{a,b\}_{ij}=a_ib_j+a_jb_i$, and $[a,b]_{ij}=a_ib_j-a_jb_i$.

We denote as $\bar\beta_i$ the structure functions which determine the hadronic tensor $H_{ij}(\xi_2)$, where
${\vec \xi_2}$ is antinucleon
polarization unit vector, corresponding to the case when the antiproton is polarized. 
The structure of this hadronic tensor is similar to the case
of polarized proton and is given by Eq. (\ref{eq:eq13}), under replacement
$\beta_i\to \bar\beta_i$.

Therefore, the dependence of the polarization observables on the
nucleon (or antinucleon) polarization is determined by 13 structure
functions. The expressions of these structure functions in terms of
the scalar amplitudes are given in the Appendix.

Let us consider the case when both initial particles are polarized.
The general structure of the hadronic tensor which describes the
correlation of the nucleon and antinucleon polarizations can be
written as
\ba
H_{ij}(\xi_1 ,\xi_2 )&=&{\vec \xi_1}\cdot {\vec m}{\vec \xi_2}\cdot {\vec m}
\Bigl (\gamma_1q_iq_j+\gamma_2m_im_j+
\gamma_3n_in_j+\gamma_4\{q,m\}_{ij}+i\gamma_5[q,m]_{ij}\Bigr )
\nn\\
&&
+{\vec \xi_1}\cdot {\vec m}{\vec \xi_2}\cdot {\vec q}
\Bigl (\gamma_6q_iq_j+\gamma_7m_im_j+
\gamma_8n_in_j+\gamma_9\{q,m\}_{ij}+i\gamma_{10}[q,m]_{ij}\Bigr )
\nn\\
&&
+{\vec \xi_1}\cdot {\vec q}{\vec \xi_2}\cdot {\vec m}
\Bigl (\gamma_{11}q_iq_j+\gamma_{12}m_im_j+
\gamma_{13}n_in_j+\gamma_{14}\{q,m\}_{ij}+i\gamma_{15}[q,m]_{ij}\Bigr )
\nn\\
&&
+{\vec \xi_1}\cdot {\vec n}{\vec \xi_2}\cdot {\vec n}
\Bigl (\gamma_{16}q_iq_j+\gamma_{17}m_im_j+
\gamma_{18}n_in_j+\gamma_{19}\{q,m\}_{ij}+i\gamma_{20}[q,m]_{ij}\Bigr )
\nn\\
&&
+{\vec \xi_1}\cdot {\vec q}{\vec \xi_2}\cdot {\vec q}
\Bigl (\gamma_{21}q_iq_j+\gamma_{22}m_im_j+
\gamma_{23}n_in_j+\gamma_{24}\{q,m\}_{ij}+i\gamma_{25}[q,m]_{ij}\Bigr )
\nn\\
&&
+{\vec \xi_1}\cdot {\vec m}{\vec \xi_2}\cdot {\vec n}
\Bigl (\gamma_{26}\{q,n\}_{ij}+\gamma_{27}\{m,n\}_{ij}+
i\gamma_{28}[q,n]_{ij}+i\gamma_{29}[m,n]_{ij}\Bigr )
\nn\\
&&
+{\vec \xi_1}\cdot {\vec n}{\vec \xi_2}\cdot {\vec m}
\Bigl (\gamma_{30}\{q,n\}_{ij}+\gamma_{31}\{m,n\}_{ij}+
i\gamma_{32}[q,n]_{ij}+i\gamma_{33}[m,n]_{ij}\Bigr )
\nn\\
&&
+{\vec \xi_1}\cdot {\vec n}{\vec \xi_2}\cdot {\vec q}
\Bigl (\gamma_{34}\{q,n\}_{ij}+\gamma_{35}\{m,n\}_{ij}+
i\gamma_{36}[q,n]_{ij}+i\gamma_{37}[m,n]_{ij}\Bigr )
\nn\\
&&
+{\vec \xi_1}\cdot {\vec q}{\vec \xi_2}\cdot {\vec n}
\Bigl (\gamma_{38}\{q,n\}_{ij}+\gamma_{39}\{m,n\}_{ij}+
i\gamma_{40}[q,n]_{ij}+i\gamma_{41}[m,n]_{ij}\Bigr ). 
\label{eq:eq14}
\ea
We see that the polarization observables,
which are due to the correlation of the nucleon and antinucleon
polarizations, are determined by 41 structure functions $\gamma_i$,
$(i=1-41)$. Their expressions in terms of the reaction scalar
amplitudes $f_i$, $(i=1-6)$ are given in the Appendix.

\section{Cross section and polarization observables}

Let us calculate the differential cross section and the 
polarization observables for an experimental setup where the pion is fully detected.

Let us consider the same experimental conditions as in Ref. \cite{AKTM}, namely: we consider an
experimental setup where the four-momentum of the pion is fully
measured and one integrates on the information given by electron-positron pair.

The phase space volume can be written as
\be
d\Gamma =\frac{d^3q_{\pi}}{2E_{\pi}}\frac{d^4q}{(2\pi
)^9}d\Gamma_q\delta^{(4)}(p_1+p_2-q-q_{\pi}),~
d\Gamma_q=\frac{d^3k_1}{2E_1}\frac{d^3k_2}{2E_2}\delta^{(4)}(q-k_1-k_2).
\label{eq:eq16}
\ee
Therefore it is necessary to perform the integration over the phase space
volume of the lepton pair. It can be easily done using the so-called
method of the invariant integration. We have to calculate the
following integral
\be
I_{\mu\nu}=\int L_{\mu\nu}d\Gamma_q, 
\label{eq:eq17}
\ee
which  depends only on the four-vector $q_{\mu}$. Taking
into account that the leptonic tensor satisfies gauge invariance: $L_{\mu\nu}q_{\mu}=L_{\mu\nu}q_{\nu}=0$, 
the integral can be written as:
\be
I_{\mu\nu}=a\left (g_{\mu\nu}-\frac{q_{\mu}q_{\nu}}{q^2}\right ),
\label{eq:eq18}
\ee
where the unknown function $a$ depends only on one variable $q^2$.
Multiplying the left- and right-hand side of this equation by the tensor
$g_{\mu\nu}$ we have
\be
3a=-4(q^2+2m^2)\int d\Gamma_q,
\label{eq:eq19}
\ee
where $m$ is the electron mass. Calculating the integral over the
lepton-pair phase space volume we obtain
\be\
a=-\frac{2\pi }{3}\beta_e(q^2+2m^2),
\label{eq:eq20}
\ee
where $\beta_e=\sqrt{1-4m^2/q^2}$.

The differential cross section can be written in terms of the contraction of the hadronic and leptonic tensors as (the averaging over the spins of the initial  particles is not included )
\be
d\sigma
=\frac{4}{I}\frac{\pi^2\alpha^2}{q^4}L_{\mu\nu}H_{\mu\nu}\frac{d^3q_{\pi}}{2E_{\pi}}
\frac{d^4q}{(2\pi )^5}d\Gamma_q\delta^{(4)}(p_1+p_2-q-q_{\pi}).
\label{eq:eq21}
\ee
Integrating over $d^4q$ and over the lepton-pair phase space volume
we obtain
\be
d\sigma =\frac{\alpha^2}{6\pi s}\frac{\beta_e
}{\beta_N}\frac{1}{q^2}\left (1+2\frac{m^2}{q^2}\right )R\frac{d^3q_{\pi}}{2\pi
E_{\pi}},
\label{eq:eq22}
\ee
where $\beta_N=\sqrt{1-4M^2/s}$ and we introduce the following
notation
\be
R=H_{\mu\nu}\left (-g_{\mu\nu}+\frac{q_{\mu}q_{\nu}}{q^2}\right).
\label{eq:eq23}
\ee
Taking into account that the hadronic tensor satisfies the following
conditions: $H_{\mu\nu}q_{\mu}=H_{\mu\nu}q_{\nu}=0$, as a
consequence of the hadronic current conservation, we can rewrite the
contraction $R$ in following form
\be
R=H_{ij}\left (\delta_{ij}-\frac{k_ik_j}{k_0^2} \right).
\label{eq:eq24}
\ee
We see that the $R$ is determined by the space components of
the hadronic tensor only.

Let us choose the following coordinate frame: the $z$ axis is
directed along the antiproton beam momentum, the virtual photon
momentum lies in the $xz$-plane and the $y$ axis is directed along
the vector ${\vec k}\times {\vec p}$. In this case the components of
the  ${\vec p}$ and  ${\vec k}$ vectors are: $p_x=p_y=0, p_z=p;$
$k_y=0, k_x=-k\sin\theta , k_z=k\cos\theta $.

In this coordinate frame the quantity $R$ can be written as
\be
R=H_{yy}+\frac{q^2+k_0^2}{2k_0^2}(H_{xx}+H_{zz})+\frac{{\vec
k}^2}{2k_0^2} \left [\cos2\theta (H_{xx}-H_{zz})+ \sin2\theta
(H_{xz}+H_{zx})\right ].
\label{eq:eq25}
\ee
In the chosen coordinate system, the different components of the hadron tensor
which enter in the expression of the cross section are
related to the functions $\alpha_i$, ($i=1-5$), by
\ba
&&H_{yy}(0)=\alpha_2, ~H_{xx}(0)+H_{zz}(0)=\alpha_1+\alpha_3,\nn\\
&&
H_{xx}(0)-H_{zz}(0)=(\alpha_3-\alpha_1)\cos2\theta
-2\sin2\theta \alpha_4,\nn\\
&&
H_{xz}(0)+H_{zx}(0)=(\alpha_3-\alpha_1)\sin2\theta +2\cos2\theta
\alpha_4.
\label{eq:eq26}
\ea
Finally, for an experimental set up where the
pion four-momentum is fully measured and integrating over the
lepton-pair variables, the unpolarized differential cross section is obtained from Eq. (\ref{eq:eq22}), replacing $R$ by $D/4$ (the factor of four is due to the average on the spins of the initial particles). $D$ corresponds to the contraction of $R$ for unpolarized leptonic and hadronic tensors and it is expressed in terms of the scalar amplitudes as: 
\be
{\cal D}=2\left [|f_1|^2+|f_2|^2+|f_3|^2+|f_4|^2+\frac{q^2}{k_0^2}(|f_5|^2+|f_6|^2)\right ].
\label{eq:eq28}
\ee
When one particle is polarized, the contraction of the leptonic tensor and the terms of the hadronic
tensor which depend on the polarization of nucleon or antinucleon,
gives
\be
R(\xi_1 )=\left (\beta_2+\beta_3+\frac{q^2}{k_0^2}\beta_1\right )\xi_{1y}.
\label{eq:eq29}
\ee
From this expression we see that only the component of the hadron polarization which is normal to the reaction plane contributes to the cross section. The two other components $\xi_{1x} $ and $\xi_{1z} $, which lie in the reaction plane, can give contribution to the cross section only if one measures the polarization of one of the leptons.

The asymmetry due to the normal component of the polarization of the proton is given by
\be
{\cal D}A_y=-4Im\left (f_1f_2^*+f_3f_4^*+\frac{q^2}{k_0^2}f_5f_6^*\right ).
\label{eq:eq30}
\ee
The asymmetry due to the polarization of the antiproton is
\be\label{eq:eq31}
{\cal D}\bar A_y=-4Im\left (f_1f_2^*-f_3f_4^*+\frac{q^2}{k_0^2}f_5f_6^*\right ).
\ee
Similarly, one obtains the following expressions for
the correlation parameters due to the polarizations of the nucleon
and antinucleon
\ba
{\cal D}C_{yy}&=&2\left [|f_1|^2+|f_2|^2-|f_3|^2-|f_4|^2+
\frac{q^2}{k_0^2}\left (|f_5|^2+|f_6|^2\right)\right ], \nn\\
{\cal D}C_{xx}&=&2\left \{ |f_4|^2-|f_3|^2-\cos2\theta
\left [|f_1|^2-|f_2|^2+\frac{q^2}{k_0^2}(|f_5|^2-|f_6|^2)\right ]+\right . \nn\\
&&\left .2\sin2\theta Re\left (f_1f_2^*+\frac{q^2}{k_0^2}f_5f_6^*\right ) \right  \} , \nn\\
{\cal D}C_{zz}&=&2 \left \{ |f_4|^2-|f_3|^2+\cos2\theta
\left [|f_1|^2-|f_2|^2+\frac{q^2}{k_0^2}\left(|f_5|^2-|f_6|^2\right)\right ]- \right . \nn\\
&&\left . 2\sin2\theta Re\left (f_1f_2^*+\frac{q^2}{k_0^2}f_5f_6^*\right )\right \}, \nn\\
{\cal D}C_{xz}&=& 2\sin2\theta
\left[|f_2|^2-|f_1|^2+\frac{q^2}{k_0^2}(|f_6|^2-|f_5|^2)\right]+4Ref_3f_4^*-
\nn\\
&& 4\cos2\theta Re\left (f_1f_2^*+\frac{q^2}{k_0^2}f_5f_6^*\right ), \nn\\
{\cal D}C_{zx}&=& 2 \sin2\theta \left  [|f_2|^2-|f_1|^2+
\frac{q^2}{k_0^2}\left(|f_6|^2-|f_5|^2\right )\right ]-4Re f_3f_4^*- \nn\\
&&4\cos2\theta Re\left(f_1f_2^*+\frac{q^2}{k_0^2}f_5f_6^*\right ), \nn\\
C_{xy}&=&C_{yx}=C_{yz}=C_{zy}=0,
\label{eq:eq33}
\ea 
where $C_{ij}$, $(i,j=x,y,z)$ are the correlation coefficients
corresponding to the $i$ component of the nucleon polarization
vector and $j$ component of the antinucleon polarization vector.

\section{Born approximation}

\begin{figure}
\mbox{\epsfxsize=15.cm\leavevmode \epsffile{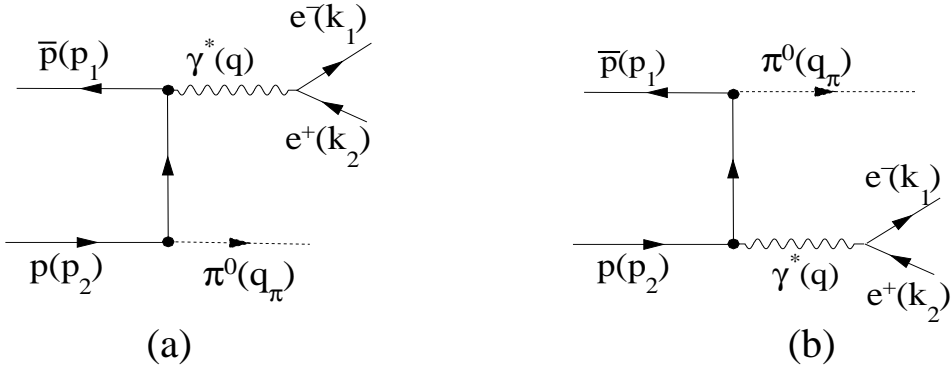}}
\caption{Feynman diagram for $\bar p+p \to e^++e^-+ \pi^0$ (a) for lepton pair emission from the antiproton and (b) from the proton.}
\label{Fig:Fig1}
\end{figure}

The numerical estimation of the differential cross section and the polarization
observables depends on the model which has been chosen to describe the $\bar
p+p\to\pi+\gamma^*$ reaction. For the present calculation of the matrix
element, let us use the approach of Ref. \cite{AKTM}, which is
based on the Compton-like Feynman amplitudes (Born approximation). The Feynman diagrams for this reaction are shown in
Fig. \ref{Fig:Fig1}a and Fig. \ref{Fig:Fig1}b for lepton pair emission from the antiproton and from the proton, respectively.
As previously discussed in Refs. \cite{DDR,AKTM}, one of the hadron involved in the electromagnetic vertexes is virtual, and, rigorously speaking, the involved FFs should be modified by taking into account off-mass-shell effects. However, we use the standard expression for the nucleon electromagnetic current involving on-mass- shell nucleons, keeping in mind that the comparison with the future data will be meaningful in the kinematical region where virtuality is small.

The following  expression for the nucleon electromagnetic current holds in terms of the Dirac $F_1(q^2)$ and Pauli $F_2(q^2)$ form factors
\be
<N(p')|\Gamma_{\mu}(q^2)|N(p)>=\bar
u(p')\left [F_1(q^2)\gamma_{\mu}+\frac{1}{4M} F_2(q^2)(\hat
q\gamma_{\mu}-\gamma_{\mu}\hat q)\right ]u(p),
\label{eq:eq31a}
\ee
where $q$ is the four-momentum of the
virtual photon. The nucleon FFs in the kinematical region
of interest for the present work are complex functions and are
largely unexplored.

Let us write the hadronic current corresponding to the two Feynman
diagrams  of Fig. \ref{Fig:Fig1} as follows
\be
J_{\mu}=\varphi_{\pi}^+(q_{\pi})\bar u(-p_1)O_{\mu}u(p_2).
\label{eq:eq32}
\ee
Here $\varphi_{\pi}$ is the pion wave function. Using the Feynman rules
we can write
\be
O_{\mu}=\frac{g}{d_1}\Gamma_{\mu}(q)(\hat q-\hat p_1+M)\gamma_5+
\frac{g}{d_2}\gamma_5(\hat p_2-\hat q+M)\Gamma_{\mu}(q),
\label{eq:eq33a}
\ee
where $d_i=q^2-2q\cdot p_i, i=1, 2, g$ is the coupling constant
describing the pion-nucleon vertex $\pi NN^*$, $N^*$ is the
off-mass-shell nucleon (the possible off-mass-shell
effects of this coupling constant are neglected). 
Note that the hadronic current (\ref{eq:eq32}) is conserved,
$q_{\mu}J_{\mu}=0$.

Let us write the matrix element in the proton-antiproton CMS. We
obtain the following expression for the amplitude ${\cal F}$ 
\be
{\cal F}=g_1{\vec \epsilon}\cdot {\vec k}{\vec \sigma}\cdot {\vec
k}+g_2{\vec \epsilon}\cdot {\vec k} {\vec \sigma}\cdot {\vec
p}+g_3{\vec \epsilon}\cdot {\vec p} {\vec \sigma}\cdot {\vec
k}+g_4{\vec \epsilon}\cdot {\vec p} {\vec \sigma}\cdot {\vec p}
  +g_5{\vec \sigma}\cdot {\vec \epsilon}
  +ig_6{\vec \epsilon}\cdot({\vec k}\times {\vec p}),
\label{34}
\ee
where $g_i$, $i=1-6,$ are the scalar amplitudes in a nonorthogonal
basis. The above structure of the matrix element arise naturally in
the transition from four- to two-component spinors.

This is a general expression which does not depend on the details of the reaction mechanism. In case of the Born approximation, which is described by the Feynman diagrams in Fig. \ref{Fig:Fig1}, the scalar amplitudes have the following form:
\ba
g_1^B&=&g\frac{p}{M}\left(\frac{1}{d_1}-\frac{1}{d_2}\right )F_2(q^2), \nn\\
g_2^B&=&\frac{g}{p}\left (\frac{1}{d_1}-\frac{1}{d_2}\right)\left[2EF_1(q^2)+\left(k_0+\frac{k}{M}p\cos\theta\right)F_2(q^2)\right ], 
\nn\\
g_3^B&=&2g\left(\frac{1}{d_1}+\frac{1}{d_2}\right )\left[\frac{p}{M}F_2(q^2)+\frac{M}{p}G_M(q^2)\right ],  
\nn\\
g_4^B&=&-2\frac{g}{p}\left( \frac{1}{d_1}+\frac{1}{d_2}\right )
\left[(2E-k_0)F_1(q^2)+\frac{k}{M}p\cos\theta F_2(q^2)\right],  
\nn\\
g_5^B&=&-2gMk\cos\theta \left(\frac{1}{d_1}+\frac{1}{d_2}\right )G_M(q^2), 
\nn\\
g_6^B&=&2g\frac{E}{p}\left(\frac{1}{d_1}+\frac{1}{d_2}\right )G_M(q^2), 
\label{eq:eq35}
\ea
where $G_M(q^2)=F_1(q^2)+F_2(q^2)$, $k$ $(p)$ is the magnitude of the virtual photon (antiproton)
momentum, and $k_0$  $ (E)$ is the energy of the virtual photon
(antiproton). All these quantities are expressed in the reaction CMS 
\be
E=\frac{\sqrt{s}}{2}, ~k_0=\frac{1}{2\sqrt{s}}(s+q^2-m_{\pi}^2),
~p^2=E ^2-M^2, ~{\vec k}^2=k_0^2-q^2,
\label{eq:eq36}
\ee
where $m_{\pi}$ is the pion mass.

The scalar amplitudes in an orthogonal basis, $f_i$ , which are used for the analysis of the polarization effects in the reaction (\ref{eq:eq1}) can be related with the scalar amplitudes $g_i$ through the following relations 
\ba
f_1&=&g_4p^2\sin^2\theta +g_5, ~f_2=(g_3k+g_4p\cos\theta
)p\sin\theta , ~f_3=g_6kp\sin\theta ,~f_4=g_5, \nn\\
~f_5&=&(g_2k+g_4p\cos\theta )p\sin\theta, 
~f_6=g_1k^2+g_5+(g_2k+g_3k+g_4p\cos\theta )p\cos\theta. 
\label{eq:eq37}
\ea
\section{Numerical results}

\begin{figure}
\mbox{\epsfxsize=10.cm\leavevmode \epsffile{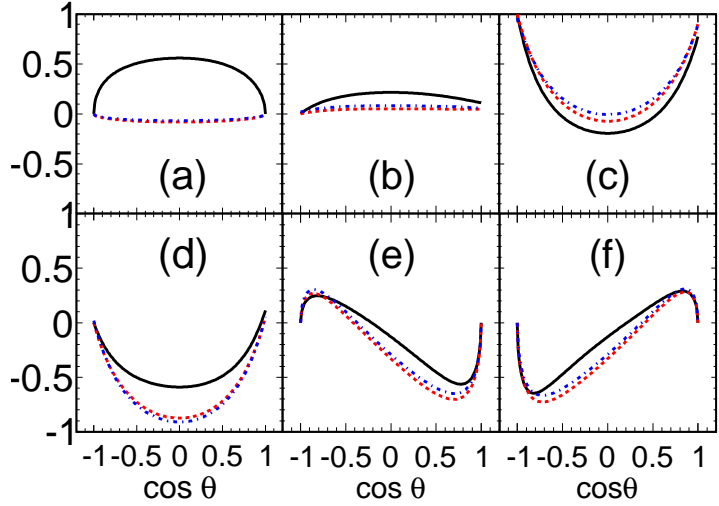}}
\caption{Polarization observables as function of $\ctp$ ($\theta$ is the angle between the momenta of the antiproton and the virtual photon) for $s=5.5$ GeV$^2$ 
and for three values of the momentum transfer squared $q^2$ : $q^2=0.5 $ (GeV/c)$^2$(solid line), $q^2= 2$ (GeV/c)$^2$(dashed line), $q^2=4$ (GeV/c)$^2$ (dash-dotted line): (a) single spin asymmetry $A_y$ ($\equiv\bar A_y$ in Born approximation); double spin correlations: (b) $C_{yy}$; (c) $C_{zz}$; (d) $C_{xx}$; (e) $C_{xz}$; (f) $C_{zx}$. 
}
\label{Fig:Fig2}
\end{figure}

\begin{figure}
\mbox{\epsfxsize=10.cm\leavevmode \epsffile{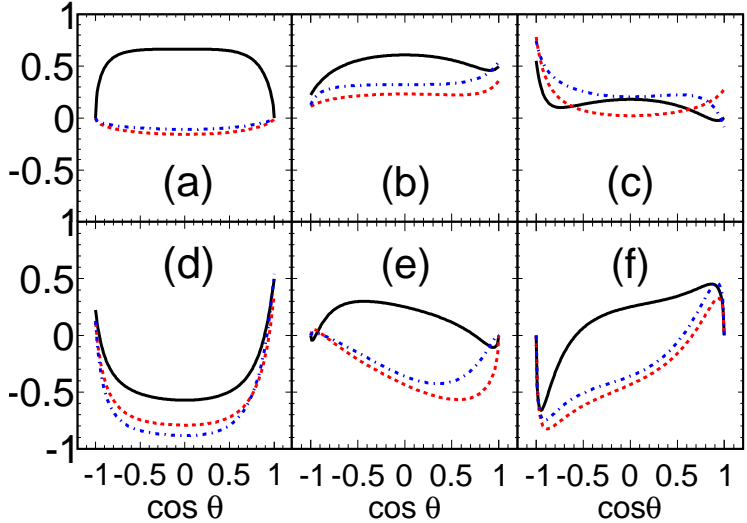}}
\caption{Same as Fig. \protect\ref{Fig:Fig2}, but for $s$=14.9 GeV$^2$ }
\label{Fig:Fig3}
\end{figure}

\begin{figure}
\mbox{\epsfxsize=10.cm\leavevmode \epsffile{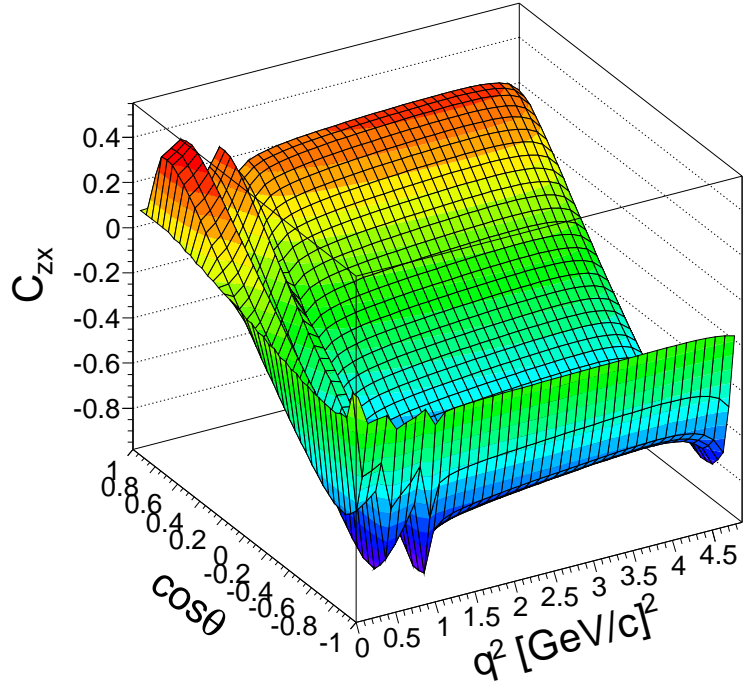}}
\caption{Bidimensional plot of $C_{zx}$ as function of $\ctp$ and $q^2$, for total energy squared $s=5.5$ GeV$^2$.}
\label{Fig:Fig4}
\end{figure}

The unpolarized cross section has been calculated in Ref. \cite{AKTM} in frame of a model based on a 'scattering channel exchange' in the Born approximation (see Fig. \ref{Fig:Fig1}). Here we focus on the numerical evaluation of the polarization observables, assuming the same mechanism. The observables are represented in Figs. \ref{Fig:Fig2}  as a function of
$\cos\theta $ for a value of the total energy squared $s$= 5.5 GeV$^2$ (corresponding to $E^{Lab}_{\bar p}=2$ GeV) and three values of $q^2$: 0.5 (GeV/c)$^2$ (solid line), 2 (GeV/c)$^2$ (dashed line) and 4 (GeV/c)$^2$ (dash-dotted line).  In Fig. \ref{Fig:Fig3}, the same quantities are plotted for $s$ = 14.9 GeV$^2$ ($E^{Lab}_{\bar p}=7$ GeV).
The single spin observables $A_y$ and $\bar{A}_y$ differ only for the sign in front of the term $f_3f_4^*$. They coincide in Born approximation, as this term is real.
A difference between these two values, if experimentally found, would be an experimental signature of the presence of terms beyond the Born approximation.
The double spin correlations are in general large and strongly depend on energy and angle. This is illustrated for a particular case in the bidimensional plot of $C_{zx}$  as function of $\ctp$ and $q^2$, Fig. \ref{Fig:Fig4}. The large structures correspond to the masses of the mesons resonances considered in the FFs parametrization from Ref. \cite{Ia73}.  
Forward and backward angles are the most favorable for the experimental measurements of this correlation, which however, vanishes in collinear kinematics. In these angular conditions, the reaction mechanism illustrated in Fig. \ref{Fig:Fig1} is also expected to be dominant \cite{By10}.
 
\section{Conclusions}
We calculated the cross section, the single spin asymmetries, and the correlation coefficients for the reaction  $\bar p+p\to \pi+\gamma^*$. We developed a model independent formalism, which allows to give simple expressions for the observables as function of the reaction amplitudes. The numerical evaluation was done assuming a 'scattering channel' mechanism, previously developed in Born approximation. We would like to stress that, however, the expressions of the observables in terms of the structure functions do not depend on the details of the reaction mechanism. The formulas listed above have a general nature as they require only P-invariance of the electromagnetic interaction,
hadron electromagnetic current conservation and the fact that the
photon has spin one.

Numerical applications in the kinematical range accessible at PANDA have been illustrated. Observables are sizable, with characteristic angular dependences.

The case when both initial hadrons are polarized and the pion is detected in
coincidence with one of the lepton, which may be polarized, will be studied in a forthcoming paper \cite{Ga10}.

The role of another possible reaction mechanism, the annihilation in 's-channel' through the exchange of a vector meson, is under investigation  \cite{By10}. Such mechanism is interesting with respect to the structure of vector mesons and to vector meson-hadron interaction. The present formalism can be applied in a straightforward way. Polarization observables  may play a crucial role in selecting the reaction mechanism, facilitating the extraction of the events which contain the information of FFs.
\section{Acknowledgments}
We are grateful to Prof. E.A. Kuraev, Dr. Yu.~M.~Bystritskiy, Dr. V.~V.~Bytev for interesting discussions and remarks. Thanks are due to Th. Hennino for a careful reading of the manuscript. This work was partly supported by  CNRS-IN2P3 (France) and by the National Academy of Sciences of Ukraine under PICS n. 5048 and by GDR n.3034 'Physique du Nucl\'eon' (France).  
\section{Appendix}

In this Appendix we give the explicit expressions for the structure functions $\alpha_i$ $(i=1-5)$,
$\beta_i$ $(i=1-13)$ and $\gamma_i$ $(i=1-41)$ in terms of the orthogonal scalar amplitudes $f_i$ $(i=1-6)$
which determine the $\bar p+p\to \pi+\gamma^*$ reaction.

The structure functions $\alpha_i$ $(i=1-5)$, describing the hadronic tensor in the case of the
unpolarized proton and antiproton, can be written as
\ba
\alpha_1&=&2[|f_5|^2+|f_6|^2],~ \alpha_2=2[|f_3|^2+|f_4|^2], ~
\alpha_3=2[|f_1|^2+|f_2|^2],\nn\\
\alpha_4&=&2Re(f_1f_5^*+f_2f_6^*), ~\alpha_5=-2Im(f_1f_5^*+f_2f_6^*). 
\label{eq:eqa1}
\ea
The expressions for the structure functions $\beta_i (i=1-13)$, describing the hadronic tensor in
the case when only proton is polarized, are
\ba
\beta_1&=&-2Imf_5f_6^*, 
~\beta_2=-2Imf_1f_2^*, 
~\beta_3=-2Imf_3f_4^*,
~\beta_4=Im(f_2f_5^*-f_1f_6^*), \nn\\ 
\beta_5&=&Re(f_2f_5^*-f_1f_6^*),~
\beta_6=-Im(f_3f_6^*+f_4f_5^*), ~
\beta_7=Im(f_1f_4^*+f_2f_3^*),\nn\\  
\beta_8&=&-Re(f_3f_6^*+f_4f_5^*), ~
\beta_9=-Re(f_1f_4^*+f_2f_3^*), ~
\beta_{10}=Im(f_4f_6^*-f_3f_5^*), \nn\\  
\beta_{11}&=&Im(f_1f_3^*-f_2f_4^*),
\beta_{12}=Re(f_4f_6^*-f_3f_5^*), ~
\beta_{13}=Re(f_2f_4^*-f_1f_3^*). 
\label{eq:eqa2}
\ea
The expressions for the structure functions $\gamma_i$ $(i=1-41)$, describing the hadronic tensor in
the case when both the proton and antiproton are polarized (spin correlations), are
\ba
\gamma_1&=&-\frac{1}{2}[|f_5|^2-|f_6|^2],~
\gamma_2=-\frac{1}{2}[|f_1|^2-|f_2|^2],~
\gamma_3=-\frac{1}{2}[|f_3|^2-|f_4|^2],~
\nn\\ 
\gamma_4&=&-\frac{1}{2}Re(f_1f_5^*-f_2f_6^*),
\gamma_5-\frac{1}{2}Im(f_2f_6^*-f_1f_5^*), 
\gamma_6=-Ref_5f_6^*,~
\gamma_7=-Ref_1f_2^*,~
\nn\\
\gamma_8&=&Ref_3f_4^*,
\gamma_9=-\frac{1}{2}Re(f_1f_6^*+f_2f_5^*),~
\gamma_{10}=\frac{1}{2}Im(f_1f_6^*+f_2f_5^*),~ 
\gamma_{11}=-Ref_5f_6^*,~
\nn\\ 
\gamma_{12}&=&-Ref_1f_2^*,~
\gamma_{13}=-Ref_3f_4^*,
\gamma_{14}=-\frac{1}{2}Re(f_1f_6^*+f_2f_5^*),~
\gamma_{15}=\frac{1}{2}Im(f_1f_6^*+f_2f_5^*), ~
\nn\\ 
\gamma_{16}&=&\frac{1}{2}[|f_5|^2+|f_6|^2],~
\gamma_{17}=\frac{1}{2}[|f_1|^2+|f_2|^2],~
\gamma_{18}=-\frac{1}{2}[|f_3|^2+|f_4|^2],~
\nn\\
\gamma_{19}&=&\frac{1}{2}Re(f_1f_5^*+f_2f_6^*),
~\gamma_{20}=\frac{1}{2}Im(f_5f_1^*+f_6f_2^*),~
\gamma_{21}=\frac{1}{2}[|f_5|^2-|f_6|^2],~
\nn\\
\gamma_{22}&=&\frac{1}{2}[|f_1|^2-|f_2|^2],~
\gamma_{23}=-\frac{1}{2}[|f_3|^2-|f_4|^2],~
\gamma_{24}=\frac{1}{2}Re(f_1f_5^*-f_2f_6^*),~
\nn\\
\gamma_{25}&=&\frac{1}{2}Im(f_5f_1^*-f_6f_2^*), 
\gamma_{26}=-\frac{1}{2}Re(f_4f_5^*+f_3f_6^*),~
\gamma_{27}=-\frac{1}{2}Re(f_1f_4^*+f_2f_3^*),~
\nn\\
\gamma_{28}&=&\frac{1}{2}Im(f_4f_5^*+f_3f_6^*),~
\gamma_{29}=-\frac{1}{2}Im(f_1f_4^*+f_2f_3^*), 
\gamma_{30}=-\frac{1}{2}Re(f_4f_5^*-f_3f_6^*),~
\nn\\
\gamma_{31}&=&-\frac{1}{2}Re(f_1f_4^*-f_2f_3^*),~
\gamma_{32}=-\frac{1}{2}Im(f_3f_6^*-f_4f_5^*),~
\gamma_{33}=-\frac{1}{2}Im(f_1f_4^*-f_2f_3^*),
\nn\\ 
\gamma_{34}&=&-\frac{1}{2}Re(f_4f_6^*+f_3f_5^*),~
\gamma_{35}=-\frac{1}{2}Re(f_1f_3^*+f_2f_4^*),~
\gamma_{36}=\frac{1}{2}Im(f_3f_5^*+f_4f_6^*),~
\nn\\
\gamma_{37}&=&-\frac{1}{2}Im(f_1f_3^*+f_2f_4^*),~ 
\gamma_{38}=-\frac{1}{2}Re(f_4f_6^*-f_3f_5^*),~
\gamma_{39}=-\frac{1}{2}Re(f_2f_4^*-f_1f_3^*),~
\nn\\ 
\gamma_{40}&=&-\frac{1}{2}Im(f_3f_5^*-f_4f_6^*),~
\gamma_{41}=-\frac{1}{2}Im(f_2f_4^*-f_1f_3^*). 
\label{eq:eqa3}
\ea
The structure of the hadronic tensor describing the polarization of  the
antiproton is the same as for the case of polarized proton. Let us
designate these structure functions as $\bar\beta_i$ and their expressions
in terms of the scalar amplitudes are
\ba
\bar\beta_1&=&-2Imf_5f_6^*, 
~\bar\beta_2=-2Imf_1f_2^*,~
\bar\beta_3=2Imf_3f_4^*,~
\bar\beta_4=Im(f_2f_5^*-f_1f_6^*), \nn\\ 
~\bar\beta_5&=&Re(f_2f_5^*-f_1f_6^*),~
~\bar\beta_6=Im(f_3f_6^*-f_4f_5^*), 
~\bar\beta_7=Im(f_1f_4^*-f_2f_3^*),
\nn\\ 
~\bar\beta_8&=&Re(f_3f_6^*-f_4f_5^*),~
\bar\beta_9=Re(f_2f_3^*-f_1f_4^*),
\bar\beta_{10}=Im(f_4f_6^*+f_3f_5^*),~
\nn\\ 
\bar\beta_{11}&=&-Im(f_1f_3^*+f_2f_4^*),~
\bar\beta_{12}=Re(f_4f_6^*+f_3f_5^*),~
\bar\beta_{13}=Re(f_2f_4^*+f_1f_3^*).
\label{eq:eqa4}
\ea


\begin{thebibliography}{99}
\bibitem{Re67}
  A.~I.~Akhiezer and M.~P.~Rekalo,
  Sov.\ Phys.\ Dokl.\  {\bf 13} (1968) 572
  [Dokl.\ Akad.\ Nauk Ser.\ Fiz.\  {\bf 180} (1968) 1081];  
  Sov.\ J.\ Part.\ Nucl.\  {\bf 4} (1974)  277
  [Fiz.\ Elem.\ Chast.\ Atom.\ Yadra {\bf 4} (1973) 662].
\bibitem{Pu10}
  A.~J.~R.~Puckett {\it et al.},
  Phys.\ Rev.\ Lett.\  {\bf 104} (2010) 242301 and Refs therein.
\bibitem{CFP08}
  C.~F.~Perdrisat, V.~Punjabi and M.~Vanderhaeghen,
  Prog.\ Part.\ Nucl.\ Phys.\  {\bf 59} (2007) 694.
\bibitem{PANDA} 
  Physics Performance Report for PANDA: Strong Interaction Studies with
  Antiprotons, The PANDA Collaboration,
  arXiv:0903.3905 [hep-ex]; 
  http://www.gsi.de/PANDA;
\bibitem{FAIR} http://www.gsi.de/FAIR. 
\bibitem{MPR}
M. P. Rekalo, Sov. J. Nucl. Phys. {\bf 1}, 760 (1965).
\bibitem{Hades} http://www-hades.gsi.de
\bibitem{DDR}
A. Z. Dubnickova, S. Dubnicka, and M. P. Rekalo, Z. Phys. {\bf C70}, 473 (1996).
\bibitem{AKTM}
C.~Adamuscin, E.~A.~Kuraev, E.~ Tomasi-Gustafsson, and F.~Maas, Phys. Rev. {\bf C75}, 045205 (2007).

\bibitem{Ad07}
  C.~Adamuscin, E.~Tomasi-Gustafsson, E.~Santopinto and R.~Bijker,
  Phys.\ Rev.\  C {\bf 78}, 035201 (2008).


\bibitem{PAX}
  P.~Lenisa and F.~Rathmann,
  CERN Cour.\  {\bf 50N6}, 21 (2010).
\bibitem{Ak77}  A.~I.~Akhiezer and M.~P.~Rekalo, {\it Hadron Electrodynamics}, Naukova Dumka, Kiev (1977).

\bibitem{Ia73}
  F.~Iachello, A.~D.~Jackson and A.~Lande,
  Phys.\ Lett.\  \textbf{ B43}, 191 (1973).
\bibitem{By10} Yu.~Bystritskiy {\it et al.}, in preparation.
\bibitem{Ga10} G.~I.~Gakh {\it et al.}, in preparation.

\end{thebibliography}
\end{document}